\documentclass[12pt]{iopart}

\pdfoutput=1

\usepackage{ifpdf}
\usepackage{iopams}  
\usepackage{multirow}

\newcommand{\JPhyG}{\em Journal of Physics G}

\newcommand{\delmot}{\mbox{$\Delta m^2_{12}$} } 
\newcommand{\delmoth}{\mbox{$\Delta m^2_{13}$} }

\newcommand{\sintot}{\mbox{$\sin^2 2\theta_{12}$} }
\newcommand{\sintoth}{\mbox{$\sin^2 2\theta_{13}$} }

\newcommand{\bnel}{\mbox{$\bar{\nu}_e$} }

\newcommand{\obb}{0\mbox{$\nu\beta\beta$ - decay} } 
\newcommand{\zbb}{2\mbox{$\nu\beta\beta$ - decay} }

\newcommand{\be}{\begin{equation}}
\newcommand{\ee}{\end{equation}}
\def\bea{\begin{eqnarray}} 
\def\eea{\end{eqnarray}} 
\newcommand{\ra}{\rightarrow }

\newcommand{\isotope}[2][]{\mbox{\ensuremath{^{#1}#2}}}

\ifpdf
	\usepackage[final,pdftex]{graphicx}
	\usepackage{epstopdf}
	\usepackage[unicode,pdfpagemode=UseNone,hyperindex=true,pdfborder={0 0 0}, plainpages=false, pdfpagelabels,breaklinks=true]{hyperref}
\else
	 \usepackage[final,dvips]{graphicx}
	\usepackage[unicode,dvipdfm,pdfpagemode=UseNone,hyperindex=true,pdfborder={0 0 0}, plainpages=false,breaklinks=true]{hyperref}
\fi

\graphicspath{{./}{./Plots/}}

\usepackage{color}
\definecolor{linkcol}{rgb}{0,0,0.4} 
\definecolor{citecol}{rgb}{0.5,0,0} 

\hypersetup
{
bookmarksopen=true,
unicode = true,
pdftitle="Solar neutrino-electron scattering as background limitation for double beta decay",
pdfauthor={N. Fiuza de Barros and  K. Zuber}, 
pdfsubject="Nuclear Physics", 
pdfkeywords={Solar Neutrinos} {Energetic Particles},
pdfmenubar=true,
pdfhighlight=/O, 
colorlinks=true,
pdfpagemode=UseNone, 
pdfpagelayout=SinglePage, 
pdffitwindow=true, 
linkcolor=linkcol,
citecolor=citecol,
urlcolor=linkcol 
}

\begin{document}
\today

\title[Solar neutrino-electron scattering as background limitation for double beta decay]{Solar neutrino-electron scattering as background limitation for double beta decay}

\author{N F de Barros, K Zuber}

\address{Institut f\"ur Kern- und Teilchenphysik, Technische Universit\"at Dresden,\\
Zellescher Weg 19, 01062 Dresden, Germany}
\ead{Nuno.Barros@tu-dresden.de, zuber@physik.tu-dresden.de}
\begin{abstract}
The background on double beta decay searches due to elastic electron scattering of solar neutrinos of all double beta emitters with Q-value larger than 2 MeV is calculated, taking into account survival probability and flux uncertainties of solar neutrinos. This work determines the background level to be $1- 2\times 10^{-7}$ counts $keV^{-1}kg^{-1}yr^{-1}$, depending on the precise Q-value of the double beta emitter. It is also shown that the background level increases dramatically if going to lower Q-values. Furthermore, studies are done for various detector systems under consideration for next generation experiments. It was found that  experiments based on loaded liquid scintillator have to expect a higher background. Within the given nuclear matrix element uncertainties any approach exploring the normal hierarchy has to face this irreducible background, which is a limitation on the minimal achievable background for purely calorimetric approaches.  Large scale liquid scintillator experiments might encounter this problem already while exploring the inverted hierarchy.
Potential caveats by using more sophisticated experimental setups are also discussed.

\end{abstract}

\submitto{\JPhyG}

\maketitle

\section{Introduction}

In the last decade the field of non-accelerator particle physics and particle astrophysics has made major progress and is now considered to be a reliable research area complementary to high energy accelerator physics.

A special subset is the field of low background physics, searching for extremely rare events normally hidden in an overwhelming large background. Best examples for this scenario are dark matter searches and neutrino physics. As part of this, it is one of the big achievements of the last decade that the observation of neutrino oscillations have shown the existence of a non-vanishing rest mass of the neutrinos. It is known that the problem of missing solar neutrinos can be explained by matter effects which is confirmed by the KamLAND reactor experiment \cite{kamland1,Araki:2004mb,Collaboration:2011uq}. Additionally, also the atmospheric neutrino anomaly observed in water \v{C}erenkov detectors like Super-Kamiokande has been confirmed by long baseline accelerator experiments \cite{Fukuda:1998mi,Adamson:2005qc}. 
All evidences can be converted into quantities of the PMNS mixing matrix elements linking weak and mass eigenstates given by \delmot, \sintot, \delmoth and  \sintoth.

However, neutrino oscillations do not fix the absolute neutrino mass scale, thus leaving two possible scenarios open: almost degenerate neutrinos, a valid scenario if the neutrino mass is larger than about 100 meV, or a hierarchical structure if the mass is smaller.
The latter can be split in a normal  ($m_1<m_2<m_3$)  or inverted ($m_3<m_1<m_2$) hierarchy, depending on the sign of the atmospheric splitting. 
In addition to beta decay end-point measurements, the process of neutrino-less double beta decay is used to explore the absolute mass scale. 
The double beta decay process, changing Z by two units while leaving the atomic number A constant can occur in two ways
\be
(Z,A) \ra (Z+2,A) + 2 e^-  + 2 \bnel \quad (\zbb)
\ee      
and
\be
(Z,A) \ra (Z+2,A) + 2 e^-  \quad (\obb) .
\ee      
The latter is of enormous importance for neutrino physics.
This total lepton number violating process is not allowed in the Standard Model and requires that neutrinos and antineutrinos are identical. Furthermore, a non-vanishing rest mass is required to account for helicity matching. The signal of this decay is a peak at the Q-value of the nuclear transition in the sum energy spectrum of the two emitted electrons. For more details see \cite{PhysRevD.25.2951,Avignone:2005cs,0954-3899-30-9-R01}. The physics quantity deduced is called effective Majorana mass and is given by

\be
\langle m_\nu \rangle
= \left| U_{e1}^2 \, m_1 + U_{e2}^2 \, m_2 \, e^{i \alpha} 
+ U_{e3}^2 \, m_3 \, e^{i\beta}\right| . 
\ee  

Here $U_{e1} = \cos \theta_{12} \, \cos \theta_{13}$, $U_{e2} = \sin
\theta_{12} \, \cos \theta_{13}$ and $U_{e3}^2 = 1 - U_{e1} ^2-
U_{e2}^2$. The current knowledge of  these mixing angles is given in table \ref{tab:oscpar}. 

\begin{table}
\caption{Neutrino mixing parameters: best-fit values as well as $1\sigma$ and 3$\sigma$ ranges \cite{Schwetz:2011qt}. Normal hierarchy was assumed.}
\footnotesize\rm
\begin{tabular*}{\textwidth}{@{}l*{15}{@{\extracolsep{0pt plus12pt}}l}}
\br
Parameter  & $\mbox{Best-fit}^{+1 \sigma}_{-1 \sigma}$ & $3 \sigma$ \\
    \mr
 $\sin^2 \theta_{12}$ & $0.316\pm 0.016$  & $0.27-0.37$\\
$\sin^2 \theta_{13}$ & $0.017^{+0.007}_{-0.009}$  & $\le 0.040$ \\ 
$\Delta m_{21}^2 \left[10^{-5}\right.$ eV$\left.^2\right]$ & $7.64^{+0.19}_{-0.18}$ & $7.12-8.23$ \\
$\Delta m_{31}^2 \left[10^{-3} \right.$eV$\left.^2\right]$ & $2.45\pm 0.09$ & $2.18-2.73$ \\
\br
\end{tabular*}
\label{tab:oscpar}
\end{table}

The life-time of \obb is inversely proportional to the effective mass squared.
\be
\left(T_{1/2}^{0\nu}\right)^{-1} = \frac{\langle m_{\nu}\rangle^{2}}{ m_{e}^{2}}G^{0\nu}\left|M^{0\nu}\right|^{2},
\label{eq:t12mnu}
\ee

where $G^{0\nu}$ is the phase space factor, including coupling constants, and $M^{0\nu}$ is the nuclear matrix element of the decay. 

Current and future experiments are aiming to improve their sensitivity into the inverted hierarchy range, which requires already hundreds of kilogram of isotopes and more and more stringent requirements on background. However, if being forced to go to ever smaller  neutrino masses, and hence longer half-lives, new background components might become important, which were not considered before. Especially dangerous are those which are irreducible.
One potential irreducible source for experiments which will be encountered are solar neutrino interactions. Evidently, the major irreducible background is \zbb which is not discussed here. In this paper we explore the background due to elastic neutrino-electron scattering caused by these neutrinos. For that we use all eleven double beta emitters with a Q-value larger than 2 MeV. 

One can also define a limit to the half-life sensitivity of an experiment, based on it's backgrounds level \cite{Avignone:2005cs}. The number of background events can be expressed as 

\be
B = b M t \Delta E
\label{eq:background}
\ee

where $b$ is the background index in  counts $keV^{-1}kg^{-1}yr^{-1}$, $M$ is the mass of the detector in $kg$, $t$ is the experiment  running time in years and $\Delta E$ is the energy resolution at the peak position of the double beta decay $Q_{\beta\beta}$. This expression is merely an approximation assuming that the background level is approximately constant in the energy resolution window $\Delta E$. As we are only considering the effect of solar neutrino background, this approximation remains valid. 

The discovery potential for a \obb experiment is usually defined as  \cite{Avignone:2005cs}

\be
N_{\beta\beta} = n_{\sigma}\sqrt{N_{\beta\beta} + B}
\label{eq:conflevel}
\ee

where $ n_{\sigma}$ is the confidence level in units of $\sigma$, $N_{\beta\beta}$ is the number of \obb events and $B$ is the number of background events. Considering a one-to-one signal to background ratio $N_{\beta\beta}\approx B$, eq.~\ref{eq:conflevel} becomes

\be 
N_{\beta\beta} = n_{\sigma}\sqrt{2 B}.
\ee

 Furthermore, the number of \obb events can also be expressed as a function of the decay rate $\lambda_{\beta\beta}$:

 \be
N_{\beta\beta} = N_{0} \lambda_{\beta\beta} t \epsilon = \frac{M \times 10^{3} \times a \times N_{A}}{W} \frac{\ln{2}}{T^{0\nu}_{1/2}}\times t \times \epsilon
\label{eq:ndecays}
 \ee

where $M$ is the detector total mass in $kg$, $N_{A}$ is the Avogadro number, $\epsilon$ is the detection efficiency, $a$ and $W$ are the isotopic abundance and the atomic weight of the \obb isotope, respectively.
Combining eq.~ \ref{eq:conflevel} and \ref{eq:ndecays} one can obtain an expression for the limit on the \obb half-life:

\be
T^{0\nu}_{1/2}(n_{\sigma}) = \frac{N_{A}\times 10^{3}\ln{2}}{\sqrt{2}n_{\sigma}} \frac{a \times \epsilon}{W}\sqrt{\frac{M \times t}{b \times \Delta E}} 
\label{eq:halflife}
\ee

\section{Background estimate}
The differential neutrino-electron scattering 
\be
\nu + e \ra \nu + e 
\ee
cross section is given by \cite{Hooft:1971ht}
\be
\frac{d \sigma}{dT} = \frac{2G_F^2m_e}{\pi}\left[g_L^2 + g_R^2\left(1-\frac{T}{E_\nu}\right)^2 - g_L g_R \frac{m_eT}{E_\nu}\right]
\ee
with $T$ as the kinetic energy of the recoiling electron, $G_F$ the Fermi constant, $g_L = \pm \frac{1}{2} + \sin^2 \theta_W$, $g_R = \sin^2 \theta_W$ and the Weinberg angle given by $\sin^2 \theta_W = 0.23116(13)$ \cite{Nakamura:2010zzi}. The sign convention for $g_L$ is + for $\nu_e$ and - for other active neutrinos. Thus the cross section is about a factor 6 higher for electron neutrinos. For the analysis, radiative corrections were taken into account \cite{Bahcall:1995fk}.

To obtain the expected electron recoil spectrum the cross section has to be folded with the corresponding solar neutrino fluxes. As all Q-values of the double beta emitters of interest are above 2 MeV only the $^8$B and hep-neutrinos have to be considered. The latter is much smaller and can be neglected, nevertheless it is taken into account for the following analysis. 
To determine the actual neutrino flux and composition for a specific isotope the survival probability $P_{ee} \equiv P_{ee} \left(\theta_{12},\theta_{13},\Delta m_{21}^{2}, \Delta m_{31}^{2}\right)$ at the specific peak energy has to be taken into account. 

\begin{figure}
\centering
\caption{Electron neutrino survival probability of \isotope[8]{B} neutrinos as a function of $E_{\nu}$ and the effect of varying the individual oscillation parameters. The black filled line represents the current best fit point\cite{Schwetz:2011qt} and the dashed lines represent the conservative $3\sigma$ variation ignoring correlation between the oscillation parameters.}
\label{fig:pee} 
\includegraphics[width=0.75\textwidth]{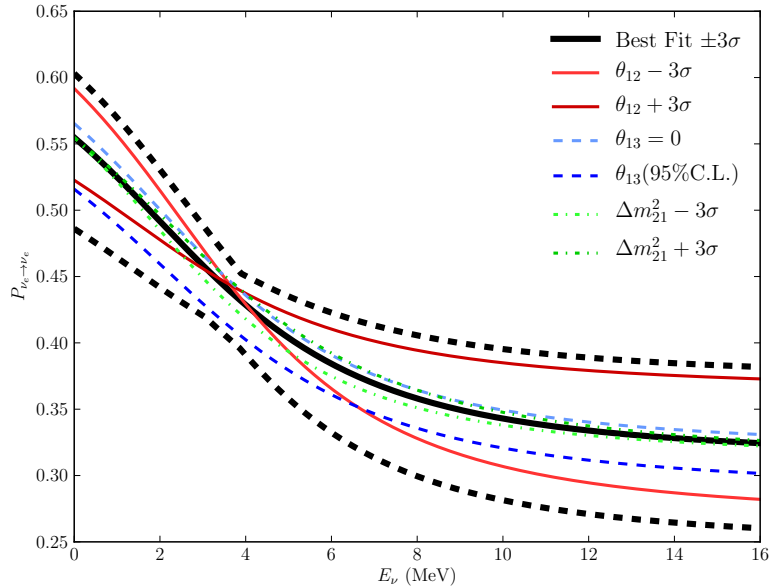}
\end{figure}

The survival probability depends on these parameters in a complex way. Figure ~\ref{fig:pee} shows the survival probability of solar neutrinos as a function of neutrino energy. In the figure, the effect of the $3\sigma$ uncertainties on each individual  parameter is also shown. It is important to notice that by varying $\theta_{12}$ inside its $3\sigma$ uncertainty effectively causes a "rotation" in the survival probability. While the $\theta_{12}+3\sigma$ results in a smaller $P_{ee}$ at lower energies and a larger one at higher energies with respect to the best fit curve, this behaviour is converted when changing to $\theta_{12}-3\sigma$. 

The effect of $\Delta m_{21}^{2}$ is small and only appears at intermediate energies (4-10 MeV), being barely noticeable in low and high energies.

In principle, in order to accurately propagate the uncertainties in the oscillation parameters, the correlation between the parameters should be taken into account. However, a more conservative approach was taken by ignoring this correlation and obtaining the maximum variation on the survival probability caused by the parameter uncertainties. Through this method, the upper and lower limit on the survival probability were obtained, which are represented as dashed black lines in figure ~\ref{fig:pee}. The most stable values of $P_{ee}$ are around 3-4 MeV.

 In order to obtain the solar neutrino induced background, the survival probability must be folded with the \isotope[8]{B} and $hep$ fluxes. As in this paper we are only considering the \obb decay candidates with a $Q_{\beta\beta}$ above 2 MeV, only these fluxes contribute to the background.
For the \isotope[8]{B} flux the measured spectrum from SNO was used\cite{paper:sno:leta}:

\be
\Phi_{\isotope[8]{B}}^{SNO} = 5.140^{+0.107}_{-0.158}(stat.)^{+0.132}_{-0.117}(syst.)\times 10^{6} cm^{-2}s^{-1} = 5.140^{+0.170}_{-0.197} \times 10^{6} cm^{-2}s^{-1}
\ee

 and for the negligible hep flux the solar model flux of \cite{bps09} is used
 
\be
\Phi_{hep} = \left(7.91\pm 1.19\right)\times 10^{3} cm^{-2}s^{-1}
\ee

\begin{figure}
\centering
\caption{Recoil energy spectrum of a single elastic scattering electron induced by a solar neutrino in the region of interest for this paper. The $3\sigma$ uncertainty band is also shown.}
\label{fig:recoilspec} 
\includegraphics[width=0.75\textwidth]{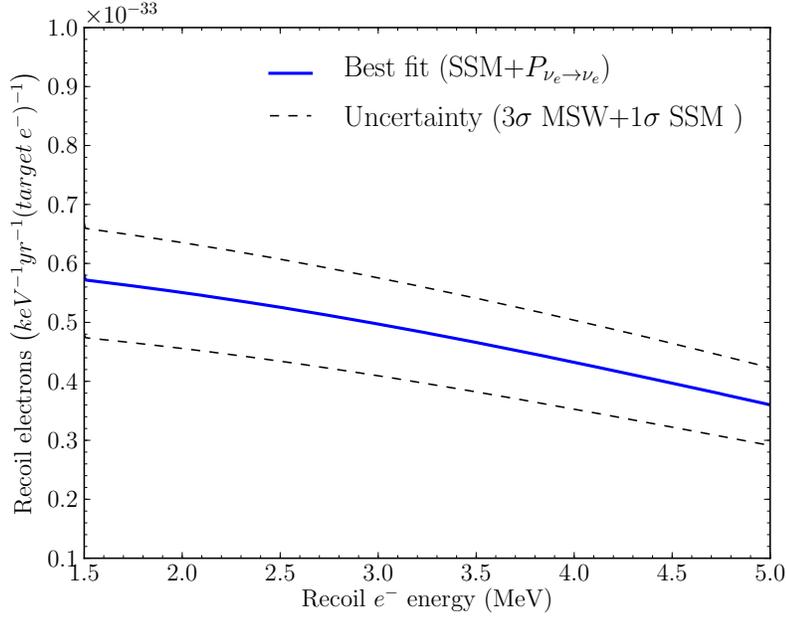}
\end{figure}

 The total resulting recoil spectrum for a single electron in the region of interest is shown in figure~\ref{fig:recoilspec}.  As mentioned, in double beta experiments only isotopes with a $Q_{\beta\beta}$ larger than 2 MeV are used , resulting in a total of 11 candidates. They are compiled in table~\ref{tab:iso}. 
 
\begin{figure}
\caption{Expected background level for each \obb candidate. It assumes an active volume composed solely of a molecular composition of the isotopes of the same species of the \obb candidate. Natural abundances and 90\% isotopical enrichment, as an example,  are assumed.}
\label{fig:bglevel}
\centering
\includegraphics[width=0.75\textwidth]{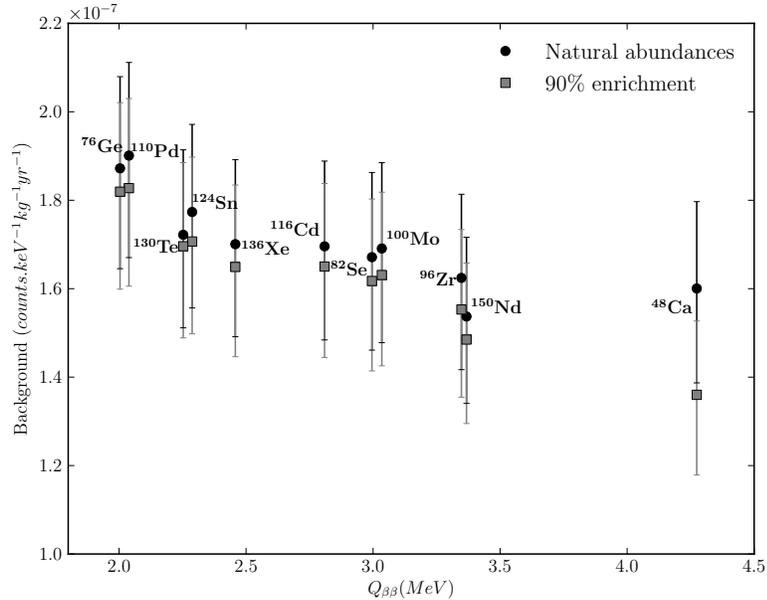}
\end{figure}

 Taking the natural abundance of the involved elements into account, the number of background events can be estimate and is shown in figure~\ref{fig:bglevel}. The numerical background levels are quoted in table~\ref{tab:background}.  All values are within 1.3 - 2.0 $\times 10^{-7}$ counts/keV/kg/yr where the highest(lowest) central value is for \isotope[76]{Ge} (\isotope[150]{Nd}). Basically three parameters determine the background level: the atomic number, i.e. the number of electrons per atom, the atomic mass, i.e. the number of atoms per kg, and the Q-value, as the number of $\nu_e$ changes as a function of neutrino energy due to flux and $P_{ee}$.  
Scaling all isotopes to a level of 90\% enrichment does not change the results dramatically, except in the case of \isotope[48]{Ca}, which, due to its low natural  abundance and small atomic mass,   enriching the double beta emitter up  to 90\% causes a 10\% decrease in the background levels. The common enrichment factor was chosen arbitrarily and can easily be adjusted to any value.

It should be noted that in these calculations only the solar neutrino induced elastic scattering backgrounds on electrons were considered. Furthermore, the background levels describe the number of recoils on a detector constructed purely from atoms which contain one of the considered \obb candidates. In fact, only a few of these candidates are planned to be studied in pure form, being most of them mixed into compounds, to facilitate the construction of a detector.  Nonetheless, it is interesting to notice that there is a clear dependence on $Q_{\beta\beta}$.

\begin{table}
\caption{\label{tab:iso} $Q$-value for the ground state to ground state transition which is calculated using isotope masses from \cite{Wapstra2003129,Audi2003337}. Values in brackets are from Penning trap measurements and the natural abundance is quoted in percent.}
\footnotesize\rm
\begin{tabular*}{\textwidth}{@{}l*{15}{@{\extracolsep{0pt plus12pt}}l}}
\br
Isotope  & Q-value $(keV)$ & nat. abund. $(\%)$\\
\mr
$^{48}$Ca	   &  4273.7 & 0.187 \\
$^{76}$Ge	  &  2039.1 (2039.04) & 7.8 \\
$^{82}$Se	    &  2995.5 & 9.2 \\
$^{96}$Zr		   &  3347.7 & 2.8 \\
$^{100}$Mo	  &  3035.0  (3034.40) & 9.6 \\
$^{110}$Pd	    &  2004.0 & 11.8 \\
$^{116}$Cd	    &  2809.1 & 7.6 \\
$^{124}$Sn	    &  2287.7 & 5.6 \\
$^{130}$Te	    &  2530.3 (2527.01) & 34.5 \\
$^{136}$Xe	   &  2461.9 (2457.83) & 8.9 \\
$^{150}$Nd	    &  3367.3 (3371.38) & 5.6 \\
\br
\end{tabular*}
\end{table}

\begin{table}
\caption{\label{tab:background} Estimated background for each \obb candidate depicted in figure~\ref{fig:bglevel}.Units are in in $($counts$.keV^{-1}.kg^{-1}.yr^{-1})$.}
\footnotesize\rm
\begin{tabular*}{\textwidth}{@{}l*{15}{@{\extracolsep{0pt plus12pt}}l}}
\hline
\obb candidate & Background (Natural ab.) ($10^{-7}$) & Background ($90\%$ enrichment) \\ 
\mr
 $^{48}Ca$   & $1.60_{-0.21}^{+0.20}$  & $1.36_{-0.18}^{+0.17}$ \\ 
 $^{76}Ge$   & $1.90_{-0.23}^{+0.21}$  & $1.83_{-0.22}^{+0.20}$ \\ 
 $^{82}Se$   & $1.67_{-0.21}^{+0.19}$  & $1.62_{-0.20}^{+0.19}$ \\ 
 $^{96}Zr$     & $1.62_{-0.21}^{+0.19}$  & $1.55_{-0.20}^{+0.18}$ \\ 
 $^{100}Mo$ & $1.69_{-0.21}^{+0.19}$  & $1.63_{-0.21}^{+0.19}$ \\ 
 $^{110}Pd$ & $1.87_{-0.23}^{+0.21}$  & $1.82_{-0.22}^{+0.20}$ \\ 
 $^{116}Cd$ & $1.70_{-0.21}^{+0.19}$  & $1.65_{-0.21}^{+0.19}$ \\ 
 $^{124}Sn$ & $1.77_{-0.22}^{+0.20}$  & $1.71_{-0.21}^{+0.19}$ \\ 
 $^{130}Te$ & $1.72_{-0.21}^{+0.19}$  & $1.70_{-0.21}^{+0.19}$ \\ 
 $^{136}Xe$ & $1.70_{-0.21}^{+0.19}$  & $1.65_{-0.20}^{+0.19}$ \\ 
 $^{150}Nd$ & $1.54_{-0.20}^{+0.18}$  & $1.49_{-0.19}^{+0.17}$ \\ 
\br
\end{tabular*}
\end{table}

\section{Further studies}

Several other features can be studied.  In this section some particular cases will be discussed.

\subsection{Low energy double beta emitters}

Besides the \obb candidates already discussed, another 24 double beta emitters with Q-values below 2 MeV exist, which are normally not considered for double beta searches because the neutrino-less decay rate scales with $Q^5$. 

\begin{figure}
\caption{Total recoil energy spectrum of elastic scattering electrons induced by solar neutrinos.  A dramatic increase in the recoil spectrum occurs at low energies where the much more intense fluxes $pp$, \isotope[7]{Be} and $pep$ dominate.}
\label{fig:totalrecoil}
\centering
\includegraphics[width=0.75\textwidth]{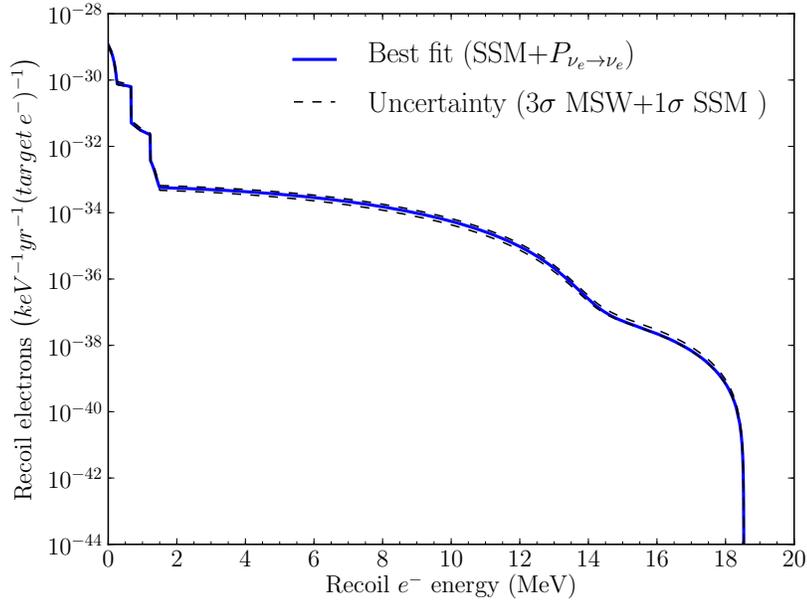}
\end{figure}

However, several experiments have some of these in their detectors. Examples would be COBRA ($^{114}$Cd), CUORE ($^{128}$Te), EXO, KamLAND-Zen
($^{134}$Xe) and SNO+ ($^{146,148}$Nd). To investigate how the background due to neutrino electron scattering changes over different energy ranges, the background of  $^{114}$Cd, with a Q-value of 534 keV, is explored as a test bench. The total recoil spectrum,  considering all solar neutrino fluxes, including CNO, for the full energy range is shown in figure~\ref{fig:totalrecoil}. 

Evidently the number of recoils increase strongly  in the low energy regime as more prominent solar flux contributions will have to be considered.  Thus, $^{114}$Cd would be effected by all solar neutrino flux components except the pp-flux. It turns out that at this energy the background would be already three orders of magnitude larger, resulting in a background rate of  $b=\left(2.09_{-0.16}^{+0.26}\right)\times 10^{-4}$ counts . $keV^{-1} kg^{-1} yr^{-1}$.

\subsection{Experimental Considerations}
\label{sec:res:compounds}

A second important consideration is closer to the experimental approaches, namely thinking about the compounds used. Most of the double beta emitters under consideration are not used as the only element in the detector, but rather in molecular compounds or dissolved. This will change the number of electrons and atoms/kg significantly and consequently  the expected background. 

We explore this effect for a series of compounds currently under consideration for use in \obb experiments: CaF$_2$ (CANDLES\cite{Yoshida:2005mm}), CdTe (COBRA\cite{Zuber:2001vm}), TeO$_2$ (CUORE\cite{Arnaboldi:2002du}), ZnSe (LUCIFER\cite{talks:lucifer}), CaMoO$_4$\cite{Belogurov:2005iq}
and liquid scintillators (LS) with a dissolved isotope like Xe (KamLAND-zen\cite{Terashima:2008zz}) or Nd (SNO+\cite{Chen:2008un}). For comparison we also show the background levels from table~\ref{tab:background} which are currently being used "in pure form": Xe (EXO\cite{Danilov:2000pp},XMASS\cite{Sekiya:2010fk}) and Ge (GERDA\cite{Abt:2004yk},MAJORANA\cite{Aalseth:2004yt}).

\begin{figure}
\caption{Expected background level for a selection of compounds  used in  \obb . For comparison, the backgrounds from \isotope{Ge} and \isotope{Xe}, which are used in pure form, are also shown. Both natural abundances and 90\% enrichment are considered.}
\label{fig:bglevel:compounds}
\centering
\includegraphics[width=0.75\textwidth]{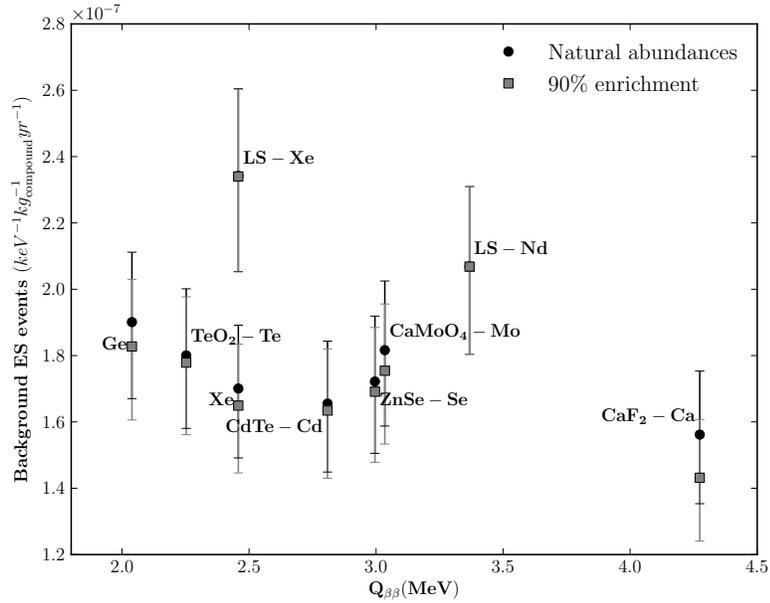}
\end{figure}

The results are shown in figure~\ref{fig:bglevel:compounds} and the numerical values are quoted in table \ref{tab:background:compound}. Like before, the study was performed both using natural abundances and 90\% double beta emitter enrichment. In this case the difference in background levels due to  the different abundances is not so pronounced, which is clearly understood by the reduced contribution of the double beta emitter to the total molecular mass and electron density. Again, the exception in this case is observed for $CaF_{2}$, which is due to the low mass of both molecular components.

\begin{table}
\caption{\label{tab:background:compound} Estimate background for each \obb candidate depicted in figure \ref{fig:bglevel}.Units in $(10^{-7} counts.keV^{-1}.kg^{-1}.yr^{-1})$.}
\footnotesize\rm
\begin{tabular*}{\textwidth}{@{}l*{15}{@{\extracolsep{0pt plus12pt}}l}}
\br
Compound & Relevant Nuclide & Background (Natural) ($10^{-7}$) & Background ($90\%$ enrichment) \\ 
\mr
LS-Xe & $^{136}Xe$                & $2.34_{-0.29}^{+0.26}$  & $2.34_{-0.29}^{+0.26}$ \\ 
LS-Nd & $^{150}Nd$                & $2.07_{-0.26}^{+0.24}$  & $2.07_{-0.26}^{+0.24}$ \\ 
$CaF_{2}$ & $^{48}Ca$          & $1.56_{-0.21}^{+0.19}$  & $1.43_{-0.19}^{+0.18}$ \\ 
$CaMoO_{4}$ & $^{100}Mo$ & $1.82_{-0.23}^{+0.21}$  & $1.75_{-0.22}^{+0.20}$ \\ 
CdTe & $^{116}Cd$                  & $1.66_{-0.21}^{+0.19}$  & $1.63_{-0.20}^{+0.19}$ \\ 
ZnSe & $^{82}Se$                     & $1.72_{-0.22}^{+0.20}$  & $1.69_{-0.21}^{+0.19}$ \\ 
$TeO_{2}$ & $^{130}Te$         & $1.80_{-0.22}^{+0.20}$  & $1.78_{-0.22}^{+0.20}$ \\ 
\mr
Pure Ge & $^{76}Ge$ & $1.90_{-0.23}^{+0.21}$  & $1.83_{-0.22}^{+0.20}$ \\ 
Pure Xe & $^{136}Xe$ & $1.70_{-0.21}^{+0.19}$  & $1.65_{-0.20}^{+0.19}$ \\ 
\br
\end{tabular*}
\end{table}

It should be noted that the units of the background level quoted in both figure~\ref{fig:bglevel:compounds} and table~\ref{tab:background:compound} are counts/keV/kg(compound)/yr. Thus, it should be kept in mind that in order to compare the sensitivity one has also to keep in mind the mass of \obb isotope present per kg of detector.

For comparison we also incorporate the results obtained from two {\em pure} compound examples: \isotope{Ge} and \isotope{Xe}. The backgrounds are in units of $(counts.keV^{-1}.kg^{-1}.yr^{-1})$. However, it is important to remark that the mass normalisation is the total detector mass. Thus, although these results give a good idea of the solar neutrino induced  background levels for these compounds, it should be kept in mind that a compound based experiment will require considerably more detector mass to have the same sensitivity as a "pure isotope" experiment. This is particularly relevant for liquid scintillator experiments (LS), whose typical detector masses are in the order of tens or hundreds of tons, while the mass of \obb isotope in these experiments is  on the order of tens or hundreds of kg.  

The results  show a steady increase when compared with the pure isotope calculations. This is not surprising, as the other atomic components of the compounds are usually lighter than the double beta emitter, increasing the density of electrons per kg. This is further confirmed by the results from $CdTe$. Tellurium, being heavier than Cd, causes a reduction of the number of electrons per kg (of detector) which translates into a lower background level than the one quoted for single Cd in table~\ref{tab:background}. 

From these results it is clear that the specific detector model used in a given experiment can intrinsically limit the achievable precision \obb. For instance, all the previous considerations describe experiments that do not have a way to directly discriminate between single a two electron events. Such a discrimination would at least partially remove the solar neutrino background and therefore improve over the achievable neutrino mass. There are under investigation several experiments that potentially could tag and therefore remove the solar neutrino background, each employing different technologies. The COBRA experiment\cite{Schwenke2011:short} has a plan to construct the detector in several pixels (sub-detectors) which would potentially allow to reconstruct the vertex of the detected events and therefore discriminate between  single and double electron events. The NEXT \cite{Collaboration:2011fk} experiment employs a similar approach consisting in a a High Pressure TPC. This will permit to reconstruct the tracks of each event providing a clear differentiation between \obb and solar neutrino events. Finally the SuperNEMO experiment\cite{Arnold:2010tu}, by using a detector where the double-beta isotope is deposited in foils also permits to independently detect each emitted electron and therefore reduce the solar neutrino background. Finally, the  EXO\cite{Flatt2007399} experiment plans to deploy a laser system to tag the daughter Barium ion resulting from the \obb. The employment of such a technology would permit to identify only double-beta decays  therefore strongly suppressing any other type of event.

\subsection{T$_{1/2}$ constraints}
\label{sec:res:t12}

Finally, the question arises  at what kind of half-life sensitivity or neutrino mass one has to start  worrying about this background. 

Fixing a measuring time of 10 years, an enrichment of 90\%, assuming 100\% detection efficiency and assuming the background discussed in this paper as the only contribution, one can  explore the half-life sensitivity as a function of the two parameters $\Delta E$ and M, by using eq.~\ref{eq:halflife}. 

\begin{figure}
\caption{$3\sigma$ bounds on $T_{1/2}$ for the considered double beta emitters as a function of active mass and energy resolution.}
\label{fig:contours:pure}
\centering
\includegraphics[width=0.95\textwidth]{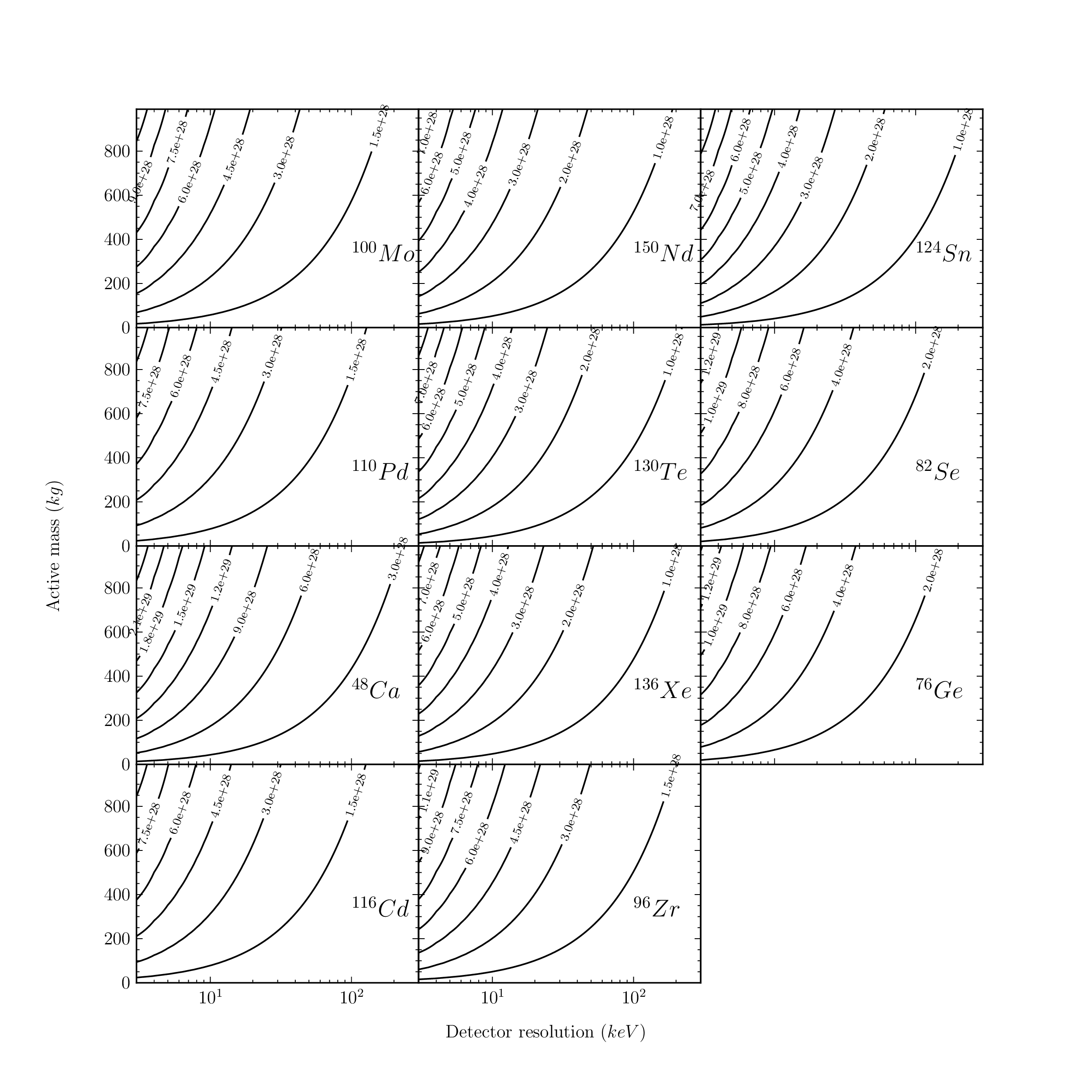}
\end{figure}

\begin{figure}
\caption{Detail of $T_{1/2}$ $3\sigma$ bounds for Ge and Xe as a function of active isotope mass and energy resolution.}
\label{fig:contours:puregexe}
\centering
\includegraphics[width=0.75\textwidth]{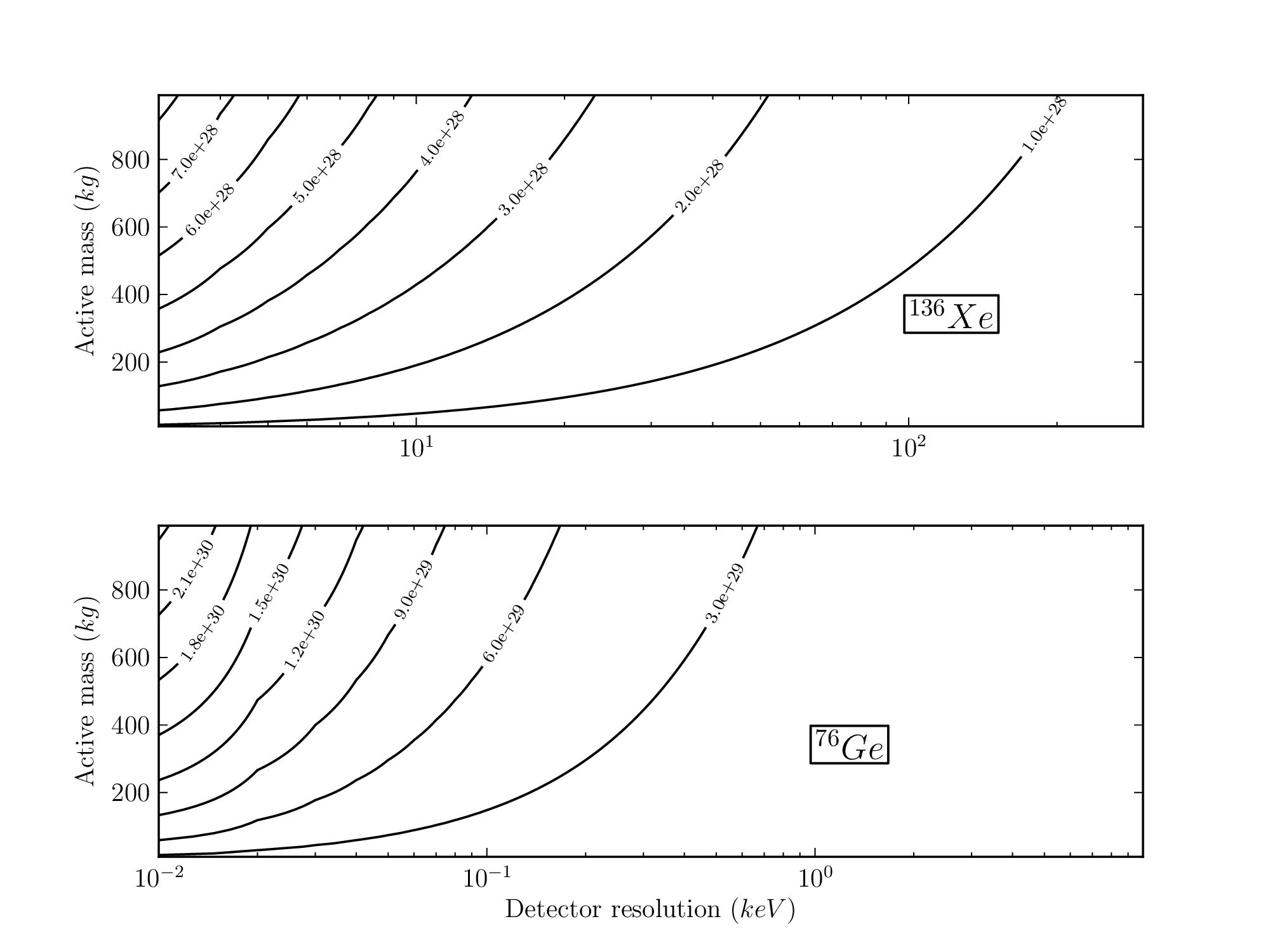}
\end{figure}

Restricting the region to 3-300 keV for energy resolution (from Ge to massively loaded scintillators) and an active mass region of 10 kg to 1t, contour plots for the half-life can be made as shown in figure~\ref{fig:contours:pure}, assuming a detector composed solely of the pure double-beta decay element. Figure~\ref{fig:contours:puregexe} shows a detail of the sensitivity for Ge and Xe, two istotopes for which there exist already experiments using them without bounding to any other element. It should be noted that in figure~\ref{fig:contours:puregexe} the range in the energy resolution for a Ge detector was zoomed to the region 0 to 10 keV, as it is unlikely to build a Ge based detector with worse resolution.

However, these results do not translate directly into the reality, as most of  these elements are planned to be used in compounds, either by bounding them with other elements, or dissolving them into some secondary compound. 

\begin{figure}
\caption{ $3\sigma$ bounds of $T_{1/2}$ for the considered double beta emitters as a function of detector mass $M$ and energy resolution $\Delta E$.}
\label{fig:contours:compounds}
\centering
\includegraphics[width=0.95\textwidth]{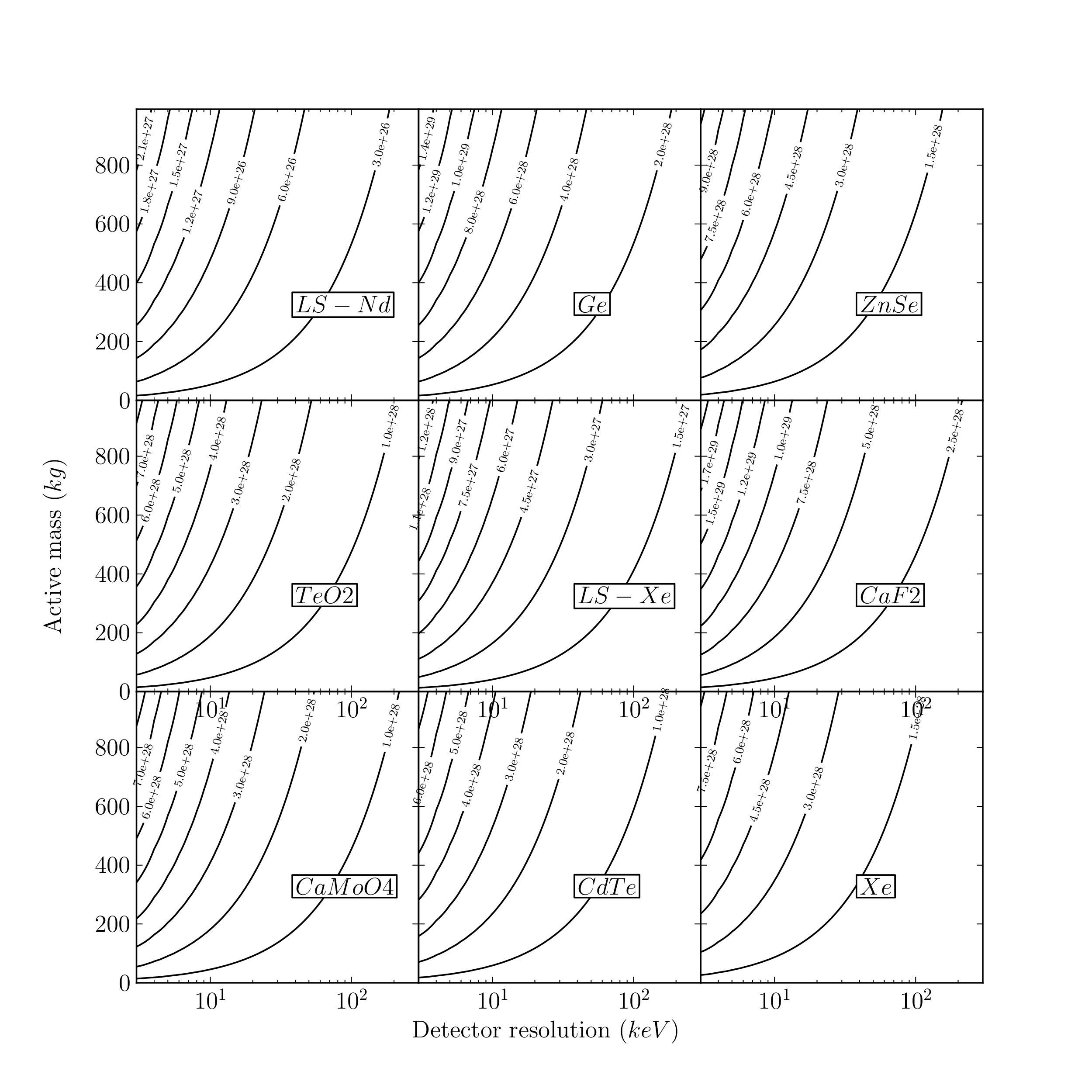}
\end{figure}

In order to better reflect the reality of present and near future experiments, this study was also performed for the compounds discussed in subsection \ref{sec:res:compounds}. The results are shown in figure~\ref{fig:contours:compounds}.

\begin{figure}
\caption{Detail of T$_{1/2}$ $3\sigma$ bounds for liquid Xe and Xe loaded liquid scintillator as a function of detector mass $M$ and energy resolution $\Delta E$.}
\label{fig:contours:compklxe}
\centering
\includegraphics[width=0.75\textwidth]{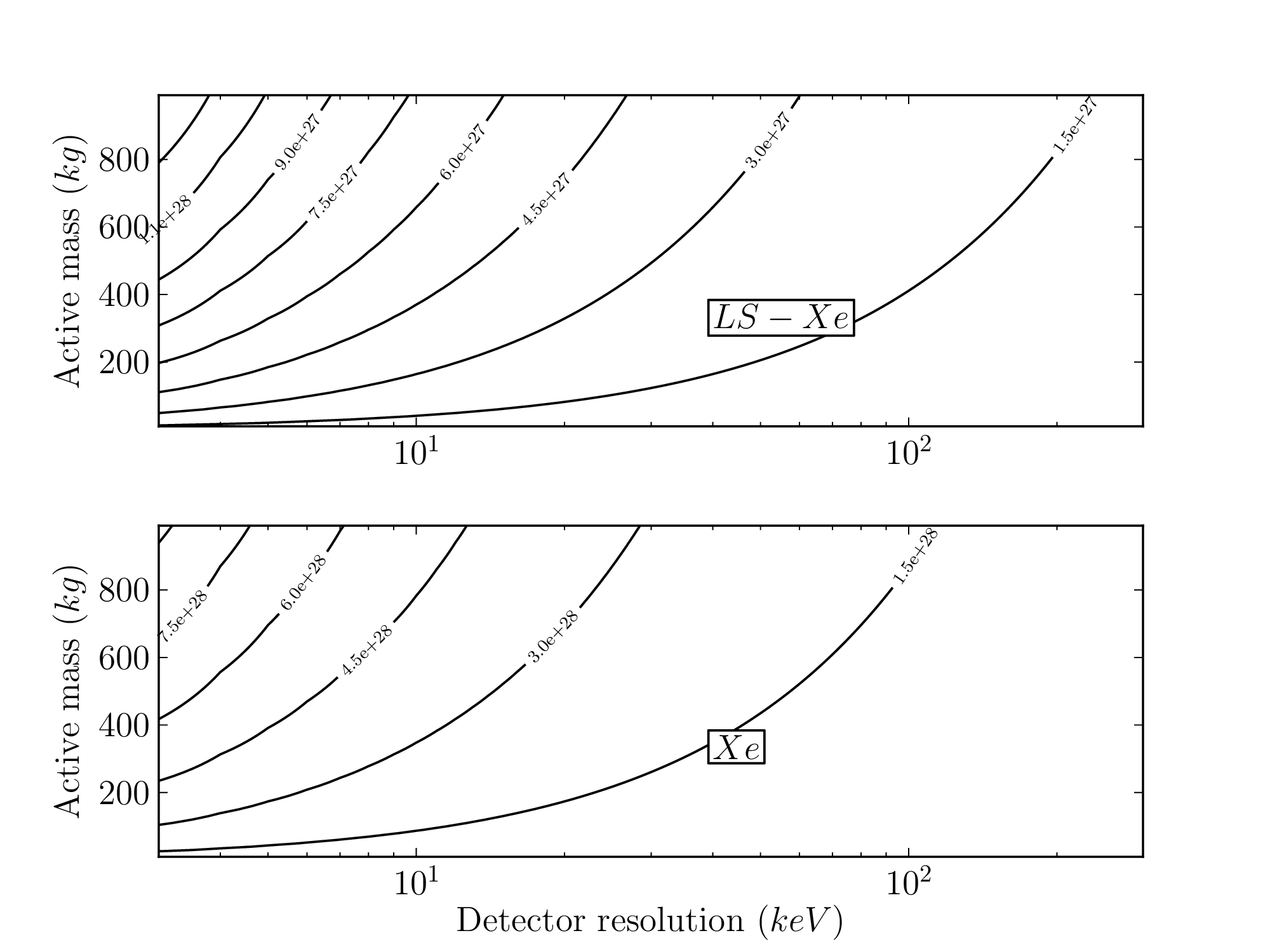}
\end{figure}

It is clearly visible that the sensitivities are considerably different due to the nature of the compounds. In figure~\ref{fig:contours:compklxe} one can see a more detailed contour map of the half-life bound for two Xe based experiments. One can see straight away that for the same total detector mass and resolution there is a difference in the half-life limit of about one order of magnitude. It should also be noted that some sensitivity is necessary when reading these figures, as the active mass does not reflect the limitations in building a detector with a large active mass. 
However, the amount of active mass considered is achievable with most detectors considered. Furthermore, it should also be taken under consideration the achievable  resolution with the particular detection techniques. For instance, liquid scintillators have worse energy resolution which also affects the limit on  T$_{1/2}$.

Using the half-life limits  shown in figure~\ref{fig:contours:pure} and figure~\ref{fig:contours:compounds}, and choosing  sensitive values for the detector mass and energy resolution,  one can establish a limit on the effective neutrino mass that can be achieved by the experiments using these compounds before worrying about solar neutrino background.  Using eq.~\ref{eq:t12mnu},  the limit on the effective neutrino mass will be:

\be
\langle m_{\nu}\rangle = \sqrt{\frac{G}{T^{0\nu}_{1/2}\left|M^{0\nu}\right|^{2}}}, \qquad\textrm{ with}\qquad G = \frac{m_{e}^{2}}{G^{0\nu}}
\ee

It should be noticed though that from this point on the results are strongly dominated by the choice of nuclear matrix elements. There are several calculations available, but most only cover a select range of double beta emitters. In this work we used the matrix elements calculated using the microscopic interacting boson model (IBM-2)\cite{Barea:2009zza,Barea:private}, except for \isotope[48]{Ca}, in which case we used the value obtained by the interacting shell model (ISM)\cite{Menendez:2008jp,Caurier:2004gf}. This was done because from IBM-2 there was a full set of calculations available for ten of the isotopes used in this study.

The phase space term used in this work was taken from \cite{Boehm:1992:nn:v2}. Table~\ref{tab:bbdata} summarises the values used for \obb candidate.

\begin{table}
\caption{\label{tab:bbdata} Phase space terms and nuclear matrix elements for the double beta decays discussed in this work. The phase space term $G^{0\nu}$   already includes the electron mass squared and has units of yr\cite{Boehm:1992:nn:v2}. The nuclear matrix element $M^{0\nu}$ is dimensionless\cite{Barea:2009zza, Menendez:2008jp}.}
\footnotesize\rm
\begin{tabular*}{\textwidth}{@{}l*{15}{@{\extracolsep{0pt plus12pt}}l}}
\br
\obb candidate & $\left(G^{0\nu}\right)^{-1} (yr)$ & $M^{0\nu} $  \\ 
\mr
\isotope[48]{Ca} & $4.10\times 10^{24}$ & 0.92 \\
\isotope[76]{Ge} & $4.09\times 10^{25}$ & 5.465 \\
\isotope[82]{Se} & $9.27\times 10^{24}$ & 4.412 \\
\isotope[96]{Zr} & $4.46\times 10^{24}$ & 2.530 \\
\isotope[100]{Mo} & $5.70\times 10^{24}$ & 3.732 \\
\isotope[110]{Pd} & $1.86\times 10^{25}$ & 3.623 \\
\isotope[116]{Cd} & $5.28\times 10^{24}$ & 2.782 \\
\isotope[124]{Sn} & $9.48\times 10^{24}$ & 3.532 \\
\isotope[130]{Te} & $5.89\times 10^{24}$ & 4.059 \\
\isotope[136]{Xe} & $5.52\times 10^{24}$ & 3.352 \\
\isotope[150]{Nd} & $1.25\times 10^{24}$ & 2.321 \\
\br
\end{tabular*}
\end{table}

 In table~\ref{tab:numassres} are the $3\sigma$ limits obtained for the $T_{1/2}$ and $\langle m_\nu\rangle$ for some selected  values, also shown in the table, of active mass and resolution, considering the detector compositions discussed in the previous sections. In these calculations it was always assumed an enrichment of 90\% of \obb isotope.
 
 From the results shown in the table, it is clearly noticeable that the solar neutrino background affect each experiment differently. While in some cases this background would only be an issue when probing the normal hierarchy region, for a few of the considered experiments this background can limit the sensitivity to the  inverted hierarchy region. Here, the liquid scintillator experiments stand out as being the most affected by this background.   
 
Several considerations should be taken into account while interpreting these results: the resolution of the detectors will strongly affect the number of background events in the analysis window. The energy resolution values were collected, where possible, from experimental publications. Some values were taken from \cite{GomezCadenas:2010gs}. Additionally, as stated before, the nuclear matrix elements vary considerably from model to model. Our choice for using the IBM-2 model was  based on the consistency of the calculation, as this model is the only one to have calculations for all isotopes discussed here, except \isotope[48]{Ca}. Any other choice of nuclear matrix elements will very likely change the achievable effective neutrino mass $\langle m_{\nu}\rangle$, but not $T_{1/2}$.

\begin{table}
\caption{\label{tab:numassres} Estimate $3\sigma$ bounds on $T_{1/2}$ and  $\langle m_\nu\rangle$ for a selection of \obb experiments over assuming 10 years of operation. }
\footnotesize\rm
\begin{tabular*}{\textwidth}{@{}l*{15}{@{\extracolsep{0pt plus12pt}}l}}
\br
Compound & \obb isotope & Mass of isotope (kg)  & $\Delta E$ (keV) & T$_{1/2}$ (yr) & $\langle m_\nu\rangle$ (meV)\\
\mr
LS-Xe			 	& \isotope[136]{Xe}   			& 500  & 230 & $1.08\times 10^{27}$ & 21.31 \\
LS-Nd 				& \isotope[150]{Nd}  				& 1000 & 220 & $2.77\times 10^{26}$ & 28.93 \\
CaF$_2$		         		& \isotope[48]{Ca}      			& 100    & 145 & $9.64\times 10^{27}$ & 22.42 \\
$CaMoO_{4}$	         		& \isotope[100]{Mo}   			& 100    & 150 & $3.81\times 10^{27}$ & 10.37 \\
CdTe				&  \isotope[116]{Cd}  			& 100    & 53 & $5.67\times 10^{28} $& 10.97 \\						
ZnSe				&  \isotope[82]{Se}     			& 100    & 5.0 & $2.65\times 10^{28} $& 4.24 \\
\multirow{2}{*}{$TeO_{2}$}& \multirow{2}{*}{ \isotope[130]{Te}}& 100   & 5.0 & $2.05\times 10^{28}$ & 4.18 \\
					&							& 500 & 5.0 & $4.58\times 10^{28} $& 2.79 \\					
\mr
Ge					&  \isotope[76]{Ge}   				& 100  & 3.5 & $4.63\times 10^{28}$ & 5.44 \\
\multirow{2}{*}{Xe}		&  \multirow{2}{*}{\isotope[136]{Xe}} & 100   & 80 & $5.69\times 10^{27} $& 9.29 \\
					& 						 	& 500   & 80 & $1.27\times 10^{28}$ & 6.22 \\
\br
\end{tabular*}
\end{table}

\section{Summary and conclusions}

We have determined the elastic neutrino-electron scattering background due to solar neutrinos for all double beta isotopes with Q$>2$ MeV. Within the current knowledge of mixing angles all values are within the range 1-2 $\times 10^{-7}$ counts/keV/kg/yr . The results show a strong dependence of the solar neutrino induced background  on the Q-value of the double-beta emitter, where the electron neutrino survival probability determines  the flavour content of the solar neutrino flux, which has a direct relation with the interaction cross sections. Furthermore, the electron density of the material used was also shown to have a strong influence in the background level, which is dominated by the atomic mass and number of the isotopes present.

It was also verified that values change significantly if actual detector systems are used. In general, it was observed that the detector systems led to a higher electron density in the detector, significantly increasing the total solar neutrino induced background. From table~\ref{tab:numassres} it was observed that this issue becomes relevant even for the inverted hierarchy regime, in the case of some experiments. In particular,  the  liquid scintillator based experiments are particularly affected, as these compounds have a large electron density and a considerably worse energy resolution.  In these cases it was observed that the neutrino mass sensitivity is almost one order of magnitude worse. 

The only way to circumvent this bound are detector systems being able to distinguish between one and two electron events, i.e. these are typically tracking detectors or to detect the daughter ion produced in double beta decay in addition to the electrons.

\ack

We thank F. Iacchello for providing us with all relevant matrix elements.
This work is supported by the Bundesministerium f\"ur Bildung und Forschung (BMBF grant 05A08OD1).

\section*{References}

\end{document}